\documentclass[preprint]{aastex}

\begin{document}

\title{The Long Wavelength Array Software Library}

\author{Jayce Dowell}
\affil{Department of Physics and Astronomy, University of New Mexico, Albuquerque, NM  87131, USA}
\email{jdowell@unm.edu}

\author{Daniel Wood}
\affil{Praxis, Inc., Alexandria, VA  22303, USA}

\author{Kevin Stovall}
\affil{Center for Gravitational Wave Astronomy and Department of Physics and Astronomy, University of Texas
at Brownsville, Brownsville, TX  78520, USA}

\author{Paul S. Ray}
\affil{U.S. Naval Research Laboratory, Washington, DC  20375 USA}

\author{Tracy Clarke}
\affil{U.S. Naval Research Laboratory, Washington, DC  20375 USA}

\and 

\author{Gregory Taylor}
\affil{Department of Physics and Astronomy, University of New Mexico, Albuquerque, NM  87131, USA}

\begin{abstract}
The Long Wavelength Array Software Library (LSL) is a Python module that provides a collection of utilities to analyze and export data collected at the first station of the Long Wavelength Array, LWA1.  Due to the nature of the data format and large-N ($\gtrsim$100 inputs) challenges faced by the LWA, currently available software packages are not suited to process the data.  Using tools provided by LSL, observers can read in the raw LWA1 data, synthesize a filter bank, and apply incoherent de-dispersion to the data.  The extensible nature of LSL also makes it an ideal tool for building data analysis pipelines and applying the methods to other low frequency arrays.
\end{abstract}

\section{Introduction}
The Long Wavelength Array \citep[LWA;][]{LWA} is a radio telescope array designed to operate in the 10 to 88 MHz frequency range.  The full array will consist of 53 stations distributed across the state of New Mexico.  The first completed station, LWA1, is designed to be a general use instrument with a diverse science case\footnote{LWA1 is funded through the National Science Foundation as a University Radio Observatory.}  Proposed projects include observations of the decametric emission from Jupiter at high temporal and spectral resolution, searches for the first stars through their influence on the HI spin temperature at redshifts of 15 to 30 ($\sim$88 to 45 MHz), and searches for emission from exoplanets and other radio transients via all-sky imaging.  For additional details of the LWA1 science case, see \citet{FL} and the references therein.

LWA1, diagramed in Figure \ref{fig:LWA1}, consists of 258 stands, each of which contains two cross-polarization dipole antennas, and an equipment shelter which houses all of the analog and digital processing electronics.  The digital processing electronics digitize the full bandwidth of LWA1 at a rate of 196 Msamples/s and have three main outputs:  a wide-band transient buffer (TBW), a narrow-band transient buffer (TBN), and a beamformer.  The TBW mode supplies a high bandwidth ($\sim$78 MHz), low duty cycle (approximately 0.03\%) data capture from all antennas.  TBN provides a continuous data stream consisting of up to 75 kHz usable bandwidth of raw voltages from all 512 antennas.  The beamformer uses delay-and-sum beamforming of all available antennas to create four independently steerable beams.  Each of these beams has two independent tunings of up to 16 MHz usable bandwidth.  It should be noted that the beamformer and one of the two transient buffers can be operated simultaneously.  Each of the three modes are streamed via user datagram protocol packets from the digital processor to five data recorder PCs which record the packet payloads on RAID storage arrays with block level striping.   

The LWA Software Library (LSL) is designed to handle the three data formats produced by LWA1 and provides basic analysis utilities.  The development of this library builds upon the LWA User's Library that was developed as part of the Long Wavelength Demonstrator Array \citep{LWDA}.  In contrast with other analysis packages, such as AIPS \citep{AIPS} and CASA \citep{CASA} which provide an end-to-end analysis environment, the overall aim of LSL is to provide the user with a set of basic data manipulation ``building blocks" that can be combined to accomplish specific tasks.  Since the primary data products are in the time domain, the focus of these ``building blocks" is on signal processing and other tasks that are typically implemented in hardware, such as filter bank synthesis and beamforming.  In keeping with this philosophy, LSL is distributed as a Python module that utilizes the object oriented nature of the Python language and supplies users of LWA1 with a structured method of interacting with their data.  In addition to avoid the high cost associated with other data analysis packages, i.e., IDL or MATLAB, using Python also allows LSL to leverage existing modules, such as NumPy \citep{Numpy}, PyEphem\footnote{\url{http://rhodesmill.org/pyephem/}}, PyFITS\footnote{PyFITS is a product of the Space Telescope Science Institute, which is operated by AURA for NASA.} \citep{PyFITS}, and AIPY \citep{AIPY}, to help extend the core functionally of the module without exposing the users to a host of new interfaces.  Thus transforming real-valued raw voltage data into the frequency domain can be accomplished with the same functions that transform complex-valued data.

Although the choice of a high-level interpreted language over a low-level compiled one such as C may appear strange for data analysis, Python provides a user-friendly interface to C libraries.  This interfacing allows performance critical functions, such as reading in the binary packed data, computing fast Fourier transforms (FFTs), and performing cross-correlations, to be written in C using optimized libraries such as ATLAS\citep{ATLAS} and FFTW \citep{FFTW} but still be accessible through the Python environment.  The added performance helps to address one of the chief challenges of LWA1 data: the data volume.  Each beam produces data at a rate of about 75 MB/s, which converts to about 264 GB of data per hour.  Since most projects will use more than one beam for calibration purposes, this potentially leads to a data rate of $\sim$1 TB/project hour.  A similar problem is faced for users of TBN where the data rate is approximately 360 GB/project hour.  The analysis of the transient buffer data is further complicated by the ``large N" aspect of these data.  A total of 33,670 baselines are possible.  In addition to this hybrid approach, the dynamic typing of variables in Python and the relative ease with which objects can be sub-classed for new purposes makes creating new analysis software more tractable.  Thus, users of LSL have the combination of high-level structure and usability for developing analysis software and low-level routines that provide enhanced performance for data manipulation.

The description of the features of LSL is broken down in this paper as follows.  The core components of the 58 LSL modules are described in \S\ref{sec:core}.  For a complete listing of all LSL modules, see Appendix \ref{sec:listing}.  Section \ref{sec:extensions} describes extensions to the core components that are available, and \S\ref{sec:examples} shows real-world applications of LSL to LWA1 data.  These demonstrate basic analysis of data from all three LWA1 modes in order to accomplish transforming data to the frequency domain, delay calibration, imaging, and analysis of beamformer data.  These examples are intended to serve as an overview of how the various components of LSL can be combined rather than a step-by-step guide to data reduction\footnote{For more detailed information about observations and data reduction, see \url{http://lwa.unm.edu/}, \url{http://www.ece.vt.edu/swe/lwa1/}, and \url{http://lwa1.freeforums.org/}}.  Section \ref{sec:obtain} details the availability of the library and concluding remarks are presented in \S\ref{sec:conclusion}.

\section{Core Components}
\label{sec:core}
There are four main components to LSL:  data readers/writers, metadata extraction routines, basic analysis routines, and general purpose routines.  Each of the components are detailed below along with their related LSL module names.

\subsection{Data Readers and Writers}
At the heart of LSL are the data readers and writers.   The readers, part of the {\tt lsl.reader} module, convert the binary packed data produced by the digital processor into Python objects that provide access to the data via NumPy arrays.  Using NumPy arrays allows the data to be manipulated with vector operations at considerable speed relative to native Python lists and enables the data to be more easily transferred into and out of the C extensions.  Due to the design of the LWA1 transient buffers, TBW and TBN data are not necessarily time ordered during the recording.  A TBN data file, for example, will have frames from various antennae interlaced in both antenna number and time.  In order to unravel the interlacing and provide time ordered data streams, LSL includes a ring buffer in {\tt lsl.reader.buffer} which accepts data frames read by the readers and returns continuous sets of data frames for further reduction.

Since LSL is not designed to be an all-inclusive analysis environment, several data writers are included to help users move their data from Python to the environment of their choice.  These writers are predominately found in the {\tt lsl.writer} module and include writers for flexible image transport system (FITS) files with tables of time series data, FITS IDI \citep{FITSIDI}, and VDIF \citep{VDIF}.  In order to ensure FITS files that are compatible with a wide range of readers, the FITS writers rely on the PyFITS module for the low-level file manipulation.  Additional writers are available via other Python modules, e.g., HDF5 with h5py\footnote{\url{http://h5py.alfven.org/}}, that support writing NumPy array data to disk.

\subsection{Metadata Extraction}
Data sets provided by LWA1 contain a variety of structured metadata, both encoded into the data streams and externally via the station's monitor and control software.  These metadata describe not only the parameters directly related to the observations, such as the pointing center, tuning frequency, and bandwidth, but also the routing of signals from the antenna to the digital processor and the overall state of the array at the time of the observation.  LSL provides the tools needed to read in the metadata from their respective sources and present the metadata to the user as Python objects.  In the case of the metadata encoded into the data stream, the readers expose the metadata as part of the object that represents the data frame.  These frame objects include methods that convert the encoded metadata values, e.g. a timestamp in digitizer samples, into more easily recognizable formats.  

Metadata generated by the monitor and control software is accessible via the {\tt lsl.common.stations} and {\tt lsl.common.metabundle} modules.  The former constructs a representation of the station wiring as a collection of objects that describe the signal path and associated system frequency responses while the latter provides access to metadata associated with observations.  The representation of the station level metadata is designed to collect and simplify it into a form that is easy to transfer between different analysis routines while maintaining the complex relationships between the various components.  At the lowest level, the unit of interest is a dipole antenna.  Thus, the station is represented as a list of antenna objects, each one containing information about the front end electronics, stand location, cables, analog receiver path, and digitizer associated with each antenna.  For any subsequent analysis the matching of a digitizer number to an antenna determines all other metadata mappings.  Similarly the observation metadata and the data use a common time source.  Thus, the timestamps associate the commands sent by the monitor and control software with their affect on the data.

\subsection{Basic Analysis Routines}
Once a data product has been read into LSL using the readers, there are a variety of routines available to analyze it to address science questions.  One of the primary differences between LWA data and the data from other telescopes is that it consists of time series voltage data as opposed to the frequency-domain data that most existing astronomical data analysis software expects.  Thus, the lowest level analysis routines are designed to transform the data into the frequency domain via FFTs or polyphase filter bank methods which are part of the {\tt lsl.correlator} module.  These methods have been generalized with a variety of options, such as overlapped FFTs, data windowing, and excising radio frequency interference (RFI) in the time domain, to make it straight forward to add additional signal processing without having to extensively modify the provided functions.  This generalization, combined with the underlying NumPy framework, also provides an avenue for other instruments to use the LSL routines.  All that is needed is for the other instruments to create a reader to load the data into a signed short integer or single precision complex NumPy array.

Since the LWA1 fundamental modes create time series data from the individual antennas, LSL provides beamfomers and an FX correlator for computing visiblities.  Beamforming is implemented in the {\tt lsl.misc.beamformer} module as both an integer sample delay-and-sum method for use with TBW data and a phase-and-sum method for TBN data.  Both methods allow the user to create an arbitrary number of pointings using the same transient buffer data without new observations in the DRX mode.  Similar to the methods described above for conversion to the frequency domain, the correlator is generalized to allow for additional signal processing and RFI excision.  Delay and polarization information are handled through the station representation created by {\tt lsl.common.stations} module.  Once correlated, the visibility data can either be saved to a FITS IDI file or be imaged with the utilities in the {\tt lsl.imaging} module.  This module uses the $w$-projection \citep{wProj} method implemented through the AIPY imaging class to deal with the wide field of view of LWA1, grid the data on the $(u,v)$ plane, and perform the inversion from the $(u,v)$ to the image plane.

In addition to conversion to the frequency domain, other functions are included to further the data analysis.  RFI identification and flagging, for example, can be accomplished by computing the spectral kurtosis statistic \citep[{\tt lsl.statistics.kurtosis};][]{SK} for the spectral data, and pulsar observations can be de-dispersed with the {\tt lsl.misc.dedispersion} module to remove the effects of the interstellar medium on the pulse arrival times.  Also included with LSL are various data simulators in the {\tt lsl.sim} module that can generate data streams to mimic those created by the digital processor.  These simulators range from functions to generate basic time-domain signals to verify other data readers to more complex functions to simulate the all-sky response of LWA1 to test imaging pipelines and source detection.  This module also provides simulations of visibility data via sub-classes of the AIPY simulation routines that are adapted to the particular case of LWA.

It should be noted that LSL has no inherent limit on the number of signals that can be processed.  The only limits to the number of antennas or baselines that can be processed at one time are set by the hardware available on the system on which the software is used.  Furthermore, the most data intensive functions, those that convert from the time to frequency domains, are written to parallelize the different input signals via the Open Multiprocessing (OpenMP) application programming interface.  This helps take advantage  of the multi-core processors that are commonly available.  Table \ref{tab:OpenMP} shows the scaling of the {\bf lsl.correlator.fx.SpecMaster} function, used for creating integrated spectra, with the number of OpenMP cores available.  Each test uses eight inputs, each with 19.6 million samples and an FFT length of 1,024 channels.  The function achieves a speed increase of approximately 3.5 by transitioning from one to four cores for both real and complex valued data, indicating that a large fraction of the code has been parallelized and that OpenMP overhead does not significantly degrade the scaling with the number of available CPU cores.  The overhead encumbered by OpenMP can also be examined by looking at the scaling of run time with the number of input signals and amount of data for a fixed number of CPU cores.  Table \ref{tab:InputScaling} shows the run times for one, two, four, eight, 16, 32, and 64, and 128 signals for real valued data and one, two, four, eight, and 16 signals for complex valued data.  In both test cases, each signal contains 19.6 million samples and an FFT length of 1,024 channels.  Also shown in this table is the effect of additional signal processing, such as time domain RFI excision and time domain data windowing, on the run time.  For the real valued data, the additional signal processing is minimal.  However, the time domain RFI excision incurs a 50 to 75\% run time penalty for the complex data due to computing the instantaneous power for each sample.  Scaling with the number of inputs is roughly linear when the number of inputs exceeds the number of CPU cores used for computation.  Below this not all compute threads are occupied and the thread synchronization can be accomplished more quickly.  Finally, the scaling with the data volume per signal is shown in Table \ref{tab:IntegrationTime} for four inputs with varying numbers of samples.  Assuming a sample rate of 19.6 Msamples/s for each test, the run times can be compared to the integration times.  In all cases, the computation time scales roughly linearly with the number of samples, indicating that the OpenMP overhead is relatively small compared to the time spent performing the FFT and integration.

\subsection{General Purpose Routines}
The LWA Software Library also includes a number of general purpose routines to supplement the LWA-specific functions.  These functions consist of a combination of mathematical routines and astronomical routines.  The mathematical routines are part of the {\tt lsl.misc.mathutil} and {\tt lsl.statistics} modules.  The first module contains functions for regridding data, Gaussian fitting, smoothing, and spherical harmonic decomposition while the latter contains robust statistical techniques based on the ROBLIB\footnote{\url{http://idlastro.gsfc.nasa.gov/ftp/contrib/freudenreich/}} library and spectral kurtosis estimators.  The astronomical routines that convert between time systems and determine source visibility can be found in the {\tt lsl.astro} and {\tt lsl.transform} modules that wrap the functions in the libnova\footnote{\url{http://libnova.sourceforge.net/}} library.  Earth orientation parameters and other geodesy information can be retrieved for a given date using the {\tt lsl.misc.geodesy} module.  In addition to the conversion functions, a version of the Global Sky Model \citep{GSM} and LFmap \citep{LFmap} via the {\tt lsl.skymap} module is available to provide information about the structure of the sky at frequencies relevant to the LWA.  Furthermore, several catalogs, e.g., the 3C catalog \citep{3C} and the ATNF pulsar catalog \citep{PSR}, are included with the package and are accessible through the {\tt lsl.catalog} module.

\section{Extensions}
\label{sec:extensions}
Although LSL itself is designed to be a general purpose library for the LWA, this does not preclude more specialized modules and analysis scripts designed for particular tasks.  These modules and scripts, collectively called extensions, use the core LSL functionality to accomplish more complex or instrument-specific tasks that are outside of the design philosophy of LSL.  In addition, the extensions are not tied to the LSL release cycle and can be updated more quickly in response to particular tasks.  The scripts included with the LSL distribution, in contrast, are meant to provide functional examples of how the various modules can be combined to accomplish general data analysis tasks.  However, like the distributed LSL scripts, the extensions are actively maintained.

 Three such extensions are already available:  {\tt Commissioning},  {\tt SessionSchedules}, and {\tt Pulsar}.  The first is a collection of commissioning scripts used at LWA1 to provide  a quick-look at data collected at the station and to aid in understanding the instrument.  For the observer, this extension provides scripts that are most similar to what is available in the AIPS or CASA environment.  For DRX data there are scripts to reduce the raw data into integrated spectra stored as either NumPy/NPZ format or HDF5 and to view the resulting file within a graphical environment where RFI can be flagged (see \S\ref{sec:drxDataExample} for an example).  The second extension is a set of scheduling tools for LWA1 observers.  These tools allow observers to define observations that are to be run at LWA1, estimate the total data volume expected, and validate observation files for both formatting and source visibility.  The final extension provides the tools needed to synthesize a filter bank and save beamformer observations to a PSRFITS file \citep{PSRFITS} utilizing the psrfits\_utils software suite\footnote{\url{https://github.com/demorest/psrfits\_utils}}.  This allows beamformer observations of pulsars to be analyzed with standard pulsar tool suites such as PRESTO \citep{PRESTO}. 

\section{Examples}
\label{sec:examples}
This section is intended to introduce some of the most commonly used tools available in LSL and to illustrate how they can be combined with other modules to complete some common analytic and scientific tasks.  It is not intended to be a cookbook but, instead,  provides an introduction to LSL while also highlighting some of the science that is possible with LWA1.

\subsection{An Introduction to LSL within the Python Environment}
Tasks in LSL can be performed in one of two ways:  either through scripts written by an observer to accomplish a particular goal or through an interactive Python session.  The interactive session is best suited for short tasks and works in a similar way to the interactive modes of other analysis software, such as MATLAB.  For example, the following can be entered into the interpreter:

\begin{verbatim}
import lsl.correlator.fx
print lsl.correlator.fx.__all__
help(lsl.correlator.fx.SpecMaster)
\end{verbatim}

The first line loads the LSL module that contains the tools to convert time series data into spectra as well as the FX correlator.  The second line prints a list of all of the functions defined in this module while the final line provides detailed help about the {\tt lsl.correlator.fx.SpecMaster()} function.  This help includes the purpose of the function, the arguments the function accepts, and any options.  When necessary, it also lists caveats associated with the function.  This help is also available on the LSL webpage at \url{http://fornax.phys.unm.edu/lwa/trac/wiki} under the ``Documentation" link.  

Similarly, a TBN data file named {\tt tbn.dat} can be read in frame by frame using:

\begin{verbatim}
import lsl.reader.tbn as tbn
fh = open('tbn.dat', 'rb')
frame = tbn.readFrame(fh)
print frame.getCentralFreq()
print frame.data.iq
\end{verbatim}

Similar to the first example, the first line loads the relevant portion of LSL into the active environment.  The second and third lines open the data file for reading and read in the first frame found in the file as a frame object, respectively.  The fourth line uses the built in metadata parsing associated with the frame object to return the tuning frequency of the TBN data.  The final line prints out the time series data associated with the frame.  Further information about the frame object and its structure can be determined by using the Python {\tt help()} function.

\subsection{Using TBW for RFI identification}
TBW data provide a brief 61 ms capture of the raw voltages at $\sim$5 ns time resolution from each of the 258 LWA1 stands.  Since this mode provides the full bandwidth available to the instrument, it is useful for identifying dipoles that have technical issues and for characterizing the RFI environment.  Since the TBW output is essentially the input to the two other LWA1 modes, these data can also be used to understand the other modes and to test more sophisticated or novelty beamforming schemes such as null steering.  The top panel of Figure \ref{fig:TBW} shows power spectral densities for stand 10 obtained  on April 18, 2012.  The process within LSL to create this figure is to read in the raw TBW data frames using the {\tt lsl.reader.tbw.readFrame()} function and then reorganize the frames into a continuous stream that contains only the data for the stand in question.  Next, the time series data must be converted to the frequency domain using the {\tt lsl.correlator.fx.SpecMaster()} function.  Finally, the spectra are plotted using the Matplotlib \citep{Matplotlib} Python module.

Panel (a) of this figure displays the bandpass and shows a typical TBW spectrum from LWA1 with 24 kHz resolution.  The shape of the spectrum is dominated by the antenna resonance point near 40 MHz and several RFI sources.  The LWA1 site is relatively free of RFI except above 88 MHz where the FM band is located and below $\sim$20 MHz where there are various Federal Communications Commission allocations for mobile land communications \citep{LWARFI}.  The strength of the RFI relative to the Galactic noise over this frequency range also demonstrates the need for 12-bit digitizers to fully represent the signals received at each dipole.  Also present in the spectra is broadband interference which is seen as low level ripples below about 60 MHz.  

The nature of this interference in the time domain is shown in panel (b) where the broadband intreference appears at bursts in digitizer levels every $\sim$ 9 ms.  This time interval corresponds to a frequency of roughly 120 Hz and the noise is likely caused by micro-sparking on a power line near the LWA1 site.  Since the power line RFI signal is confined to a relatively brief amount of time, the RFI can be easily excised by dropping the FFT windows where the value of any sample within the window exceeds a certain value.  This excision method is built into the {\tt lsl.correlator.fx.SpecMaster()} function via the {\tt ClipLevel} keyword that controls when FFT windows should be dropped.  The lower panel in Figure \ref{fig:TBW} shows the results of this excision procedure applied using a clip level of 250.  The spectra are noticeably cleaner and the strength of the ripples is reduced.

Beyond simply removing the power line RFI the data can also be used to help localize the part of the power line where the micro-sparking is occurring.  The high time resolution of the data combined with the cable delays makes it possible to calculate the arrival times of the burst at each stand in the array.  The arrival times along with the stand positions can be used to perform multilateration to yield a position for the source.  For the data presented here, the localization favors a source located north east of the LWA1 site.  This line-of-sight contains several potential sources of RFI, including two power lines and the Very Large Array visitor's center.

\subsection{Delay Calibration with TBN}
In order to create well formed beams with delay-and-sum beam forming, we need to know the total system delays for each antenna to within one or two digitizer samples (approximately 5 to 10 ns).  Although {\it a priori} knowledge of cable lengths captures the bulk of this delay, variations in the system due to temperature changes and stretching of the cables during installation lead to uncertainties.   At high frequencies, the delay can be determined by observing an isolated, bright calibration source and cross-correlating the voltages from each antenna to an antenna with known cable delay.  The phase structure as a function of frequency then provides the delay difference between the two cables after the geometric delay is removed and the unknown delays can be computed.

The delay calibration is more complicated at the low frequencies of LWA1.  First, each antenna has a wide field of view that contains multiple bright sources that contaminate the simple single calibration source picture.    Thus, the voltage measured at each antenna is a combination of all of the visible sources.  Consequently, a simple correction for the geometric delay cannot be used.  Second, the relatively small size of the station ($\sim$100 m) and the proximity of antennas to each other leads to correlated sky noise \citep{SkyNoise}.  The correlated sky appears at a fringe rate of zero and is present at all times.  Finally, the diffuse emission from the Galactic plane further complicates the phase structure and has an increasing importance towards lower frequencies.  

The solution to these problems is to isolate a particular calibration source using fringe stopping and to use a reference antenna that is located a sufficient distance from the station to resolve out the diffuse emission.  Fringe stopping compensates for the change in delay as the target source moves across the sky and stops the phase rotation of the target's fringes.  In contrast, other sources pick up additional phase rotation that causes their time averaged contribution to zero out.  Thus, this technique is effective at suppressing both contaminating sources as well as the sky noise provided that the target source is well separated in fringe rate space from each contaminant \citep{TBNFun}.  In practice, the ideal data for calibration is collected with TBN near the Cyg A transit.  At transit the fringe rate of Cyg A is well separated from Cas A and the correlated sky.   In order for fringe stopping to be effective, the observations need to span $\sim$1.5 Cyg A fringe rotation periods.  TBN data provides a continuous stream of raw voltages from all antennas but with a limited bandwidth.  To overcome this, multiple TBN captures are needed around transit to build up the wide frequency coverage required to unwrap the phase and determine the delays.

After collecting the data, it can be read in using the {\tt lsl.reader.tbn.readFrame()} function and reordered in time with the {\tt lsl.reader.buffer.TBNFrameBuffer} class.  The reordering is necessary to reassemble the data into a stream that is continuous in time for each antenna.  Once the data have been read in, the signals can be cross-correlated with the reference antennas.  The {\tt lsl.common.stations.lwa1.getAntennas()} provides the metadata necessary to relate the frame IDs found in the data to the actual antennas and their physical locations.  Fringe stopping is then applied to the visibility data.  First, the fringe rate of Cyg A is calculated for the time of the observation using the previously obtained metadata and then this rate is used to shift Cyg A to the DC component.  The contaminating sources and the correlated sky pick up a time variation and average out over the 1.5 Cyg A fringe rotation periods.  The geometric delay is also removed in this stage such that all of the phase structure remaining in the data is related to the cable delays.  Finally, the fringe stopped data are combined across multiple frequencies to get the phase as a function of frequency and, hence, the difference in cable delays.

\subsection{All-Sky Imaging with TBN}
A Stokes I image of the sky visible to LWA1 at 74.03MHz is shown in Figure \ref{fig:TBN} as a Mollweide projection of a HEALpix \citep{HEALpix} map.  No flux calibration has been applied to the map.  The map is a composite of 10 second integrations collected every 30 minutes over a 23 hour period.  Each individual snapshot consists of data from 225 stands and contains 25,200 baselines.  The various snapshots of the sky are correlated using the {\tt lsl.correlator.fx.FXMaster()} function which uses the station-level metadata returned by the {\tt lsl.common.stations.lwa1} station representation to compute both the cable delays for each stand as well as the geometric delay for a zenith pointing.  After correlating the signals and integrating for 10 seconds, the array geometry and visibility data are saved to a FITS IDI file via the interface provided by the {\tt lsl.writer.fitsidi.IDI} class.

Once all of the snapshots have been written to FITS IDI files, the data are read in and combined into the HEALpix map.  First, the {\tt lsl.imaging.utils.CorrelatedData} class is used to load in the visibility data, antennas used in the data, and the time of observation.  The visibilites are then gridded using the {\tt lsl.imaging.utils.buildGriddedImage()} function.  This function performs a $w$-projection on the data as it is gridded onto the $(u,v)$ plane using the AIPY imaging utilities.  After gridding and inversion to the image plane, the images are corrected for the primary beam response using the beam parameterization that is included with LSL.  The primary beam corrected data are then up-sampled and placed on a HEALpix map using the healpy{\footnote{\url{https://github.com/healpy/healpy}} Python module.  The up-sampling is needed to help deal with projection issues that arise at low sky elevations.  Data where the primary beam sensitivity is less than 10\% are excluded from the map in order to help mask sources of RFI around the horizon.  After all of the snap shots have been projected the final map is downgraded to a resolution of $\sim$1$^\circ$.

The Galactic plane, Cyg A, Cas A, Tau A, and the Sun are visible in the resulting map.  The striations at lower declinations are likely the result of the course sampling of the sky.  Also visible in the image are two patches of emission located to the upper right and lower left of the Galactic plane which correspond to side lobes of the plane.  

\subsection{Analysis of DRX Data}
\label{sec:drxDataExample}
A 20 minute DRX (beam former) observation centered at 30 MHz of Jupiter during a decametric burst is shown in Figure \ref{fig:DRX}.  Since beamformer data are provided to observers as raw voltages, the {\tt Commissioning} extension to LSL provides a variety of tools to convert the data into spectra.  In this example the {\tt hdfWaterfall.py} script was use to take the raw data and generate a waterfall plot with 8,192 channels ($\sim$ 2.4 kHz/channel) and five second time resolution.   This script uses the {\tt lsl.reader.drx} module to read in the raw data and the {\tt lsl.correlator.fx.SpecMaster()} function to perform the FFTs and integration.  The HDF5 files created by {\tt hdfWaterfall.py} can be interactively viewed using the {\tt plotHDF.py} script or loaded into a variety ofj other analysis environments such as MATLAB.  In the case of {\tt plotHDF.py},  the window is divided into three main sections:  a waterfall diagram on the upper left, the total power for the inner 75\% of the band as a function of time on the upper right, and the spectrum for a particular integration on the bottom.  The spectrum is selected by clicking on the waterfall display.  The viewer also allows for flagging of the waterfall in frequency and time via the keyboard and mouse. 

The burst shown in Figure \ref{fig:DRX} was recorded on December 22, 2011 starting at 01:11:55 UTC when the Io-phase is 188$^\circ$ and the central meridian longitude was 232$^\circ$.  Based on the probability distribution determined from long-term low-frequency observations of Jupiter \citep{Bigg} this event is anticipated to be a an non-Io-A event.  The emission shows bursts that have a complex time and frequency structure.  Faraday lanes consisting of bright bands of emission that are nearly constant in frequency are clearly visible in the single X polarization data shown.  These bands are a result of rotation of the linearly polarized component of the Jovian decametric emission undergoing Faraday rotation \citep{Warwick} while passing through the magnetized, ionized medium between the emission point and the LWA1 receiver.  The majority of the Faraday rotation is generated within the Earth's ionosphere.  The spectrum displayed on the bottom of the figure provides another view of the frequency structure of the burst.  The overall shape of this spectrum is dominated by the LWA1 analog bandpass filter used for this observation which is designed to suppress RFI below 20 MHz.  Once the bursts have been identified at relatively low time resolution, the data corresponding to the bursts can be re-reduced at a higher temporal resolution to provide additional information on the structure of the bursts.

\section{Availability}
\label{sec:obtain}
LSL and the three extensions mentioned in Section \ref{sec:extensions} are available under version two of the the GNU Public License for Linux, FreeBSD, and Mac OSX platforms at \url{http://fornax.phys.unm.edu/lwa/trac/wiki}.  This webpage also features a complete listing of the functions available in LSL, the current status of development of the module, and a ticket system for reporting bugs and requesting new features.  This webpage also provides a collection of short tutorials and an interface to the LSL subversion repository where the current development version can be found.

\section{Conclusion}
\label{sec:conclusion}
The capabilities of LWA1 offer a new window into the radio sky at low frequencies where a variety of science questions remain unanswered.  The data created by this instrument also provide several data analysis challenges that are not easily met with existing software and need new analysis techniques.  The LWA Software Library provides a Python module to interact with these data in an open environment where such tools can be rapidly developed and tested.  In particular, LSL provides a collection of signal processing tools oriented for astronomical data that are not readily available in existing data analysis packages.  These tools are also designed without any {\it a priori} assumptions about the scale or nature of the data by utilizing dynamic typing available in Python and the relative ease of passing processor intensive tasks to a parallelized C extension.  Thus, processing 512 signals of 12-bit real-valued voltage data is accomplished with the same function that processes 4 signals of 4-bit complex-valued voltage data (DRX).  The library also provides tools for investigating new methods, such as null-steering beamforming, on existing data without having to invest significant time developing the fundamental software from scratch.

\acknowledgements{The authors acknowledge the helpful comments from the anonymous referee.  Construction of the LWA has been supported by the Office of Naval Research under Contract N00014-07-C-0147. Support for operations and continuing development of the LWA1 is provided by the National Science Foundation under grant AST-1139974 of the University Radio Observatory program.}

\appendix
\section{Listing of LSL Modules}
\label{sec:listing}
Here we provide an alphabetical listing of the various modules and submodules within LSL.  For each module, we provide a brief description of the purpose of the module.

\begin{itemize}
	\item {\bf lsl.astro} - Astronomical utility functions and classes based on the libnova library.
	\item {\bf lsl.astro\_array} - Array-based astronomical calculations implemented in C.
	\item {\bf lsl.catalog} - Astronomical source catalogs.
	
	\item {\bf lsl.common}
	\begin{itemize}
		\item {\bf lsl.common.constants} - Commonly used physical constants.
		\item {\bf lsl.common.dp} - Information about the digital filters and beam forming implemented in the LWA1 digital processor.
		\item {\bf lsl.common.mcs} - Utilities for decoding monitor and control system metadata codes.
		\item {\bf lsl.common.metabundle} - Functions for working with observational metadata.
		\item {\bf lsl.common.paths} - Listing of data and test paths within LSL.
		\item {\bf lsl.common.progress} - Implementation of a console progress bar.
		\item {\bf lsl.common.sdf} - Functions for parsing and representing LWA1 observation specification files.
		\item {\bf lsl.common.sdm} - Utilities to parse dynamic station-level metadata.
		\item {\bf lsl.common.stations} - Structured object representations of LWA1.
		\item {\bf lsl.common.warns} - Module for generating import warnings for experimental modules.
	\end{itemize}
	
	\item {\bf lsl.correlator}
	\begin{itemize}
		\item {\bf lsl.correlator.filterbank} - Implementation of a uniform discrete Fourier transform filter bank.
		\item {\bf lsl.correlator.fx} - Routines for channelization and cross-correlation of raw voltage data.
		\item {\bf lsl.correlator.uvUtils} - Function to deal with calculations of $(u,v,w)$ coordinates and tracks.
		\item {\bf lsl.correlator.visUtils} - Tools for visualizing visibility data.
	\end{itemize}
	
	\item {\bf lsl.imaging}
	\begin{itemize}
		\item {\bf lsl.imaging.deconv} - Prototype deconvolution module for all-sky images.
		\item {\bf lsl.imaging.utils} - Utilities for reading in visibility data from FITS IDI files and forming images.
	\end{itemize}
	
	\item {\bf lsl.libnova} - Wrapper around the libnova functions.
	
	\item {\bf lsl.misc}
	\begin{itemize}
		\item {\bf lsl.misc.OrderedDict} - Backport of the Python 2.7 {\bf OrderedDict} class.
		\item {\bf lsl.misc.autostereogram} - Functions for converting two dimensional array data to autostereograms.
		\item {\bf lsl.misc.beamformer} - Implementations of integer delay-and-sum and phase-and-sum beamforming for raw voltage data.
		\item {\bf lsl.misc.dedispersion} - Functions for both incoherent and coherent dedispersion of raw voltage data.
		\item {\bf lsl.misc.geodesy} - Module to retrieve Earth orientation parameters for a given date.
		\item {\bf lsl.misc.mathutil} - Collection of mathematical utilities for smoothing and fitting data.
\end{itemize}

	\item {\bf lsl.reader}
	\begin{itemize}
		\item {\bf lsl.reader.buffer} - Ring buffer for reassembling an interleaved TBN data stream.
		\item {\bf lsl.reader.drspec} - Module for working with binary packed data from the experimental LWA1 data recorder spectrometer mode.
		\item {\bf lsl.reader.drsu} - Functions to provide direct access to LWA1 data stored on the data recorder RAID storage arrays.
		\item {\bf lsl.reader.drx} - Reader for the binary packed DRX data format.
		\item {\bf lsl.reader.errors} - Exceptions to handle errors from the the various readers.
		\item {\bf lsl.reader.tbn} - Reader for the binary packed TBN data format.
		\item {\bf lsl.reader.tbw} - Reader for the binary packed TBW data format.
	\end{itemize}
	
	\item {\bf lsl.sim}
	\begin{itemize}
		\item {\bf lsl.sim.dp} - Function for simulating raw voltage data streams.
		\item {\bf lsl.sim.drx} - Module for creating properly formatted DRX data packets.
		\item {\bf lsl.sim.errors} - Exceptions to handle errors from the various simulation modules.
		\item {\bf lsl.sim.nec\_util} - Utilities for analyzing electromechanical simulations on antenna response patterns.
		\item {\bf lsl.sim.tbn} - Functions for creating properly formatted TBN data packets.
		\item {\bf lsl.sim.tbw} - Functions for creating properly formatted TBW data packets.
		\item {\bf lsl.sim.vis} - Utilities for simulating visibility data.
	\end{itemize}
	
	\item {\bf lsl.skymap} - Classes and methods to model sky brightness and visibility.
	
	\item {\bf lsl.statistics}
	\begin{itemize}
		\item {\bf lsl.statistics.kurtosis} - Implementation of spectral kurtosis.
		\item {\bf lsl.statistics.robust} - Robust statistical methods based on ROBLIB.
		\item {\bf lsl.statistics.stattests} - A collection of statistical tests not found in any of the common
Python libraries.
	\end{itemize}
	
	\item {\bf lsl.transform} - Methods for time and position transformations.
	\item {\bf lsl.version} - Version information for LSL.
	
	\item {\bf lsl.writer}
	\begin{itemize}
		\item {\bf lsl.writer.fitsidi} - Writer for FITS interferometry data interchange convention.
		\item {\bf lsl.writer.sdfits} - Writer for the single dish FITS convention.
		\item {\bf lsl.writer.tsfits} - Writer for storing raw voltage data in FITS binary tables.
		\item {\bf lsl.writer.vidf} - Writer for the VLBI Data Interchange Format.
	\end{itemize}
\end{itemize}

\clearpage
\begin{deluxetable}{ccccc}
\tablecolumns{5}
\tablewidth{0pc}

\tablecaption{Run time scaling of the {\bf lsl.correlator.fx.SpecMaster} function with number of CPU cores\label{tab:OpenMP}}

\tablehead{
	\colhead{Cores} & \multicolumn{2}{c}{Run Time\tablenotemark{a} (s)} & \multicolumn{2}{c}{Speedup} \\
	\cline{2-3}  \cline{4-5}
	~ & \colhead{Real} & \colhead{Complex} & \colhead{Real} & \colhead{Complex}
}

\startdata
1 & 3.27 & 4.04 & 1.00 & 1.00 \\
2 & 1.71 & 2.13 & 1.88 & 1.90 \\
3 & 1.36 & 1.67 & 2.36 & 2.42 \\
4 & 0.93 & 1.16 & 3.45 & 3.48 \\
\enddata

\tablenotetext{a}{Computations were run on an Intel i7 quadcore CPU clocked at 2.93 GHz with 16 GB of RAM.}

\end{deluxetable}

\clearpage
\begin{deluxetable}{ccccccccc}
\tablecolumns{9}
\tablewidth{0pc}

\rotate
\tabletypesize{\footnotesize}

\tablecaption{Run time scaling of the {\bf lsl.correlator.fx.SpecMaster} function with number of inputs.\label{tab:InputScaling}}

\tablehead{
	\colhead{Inputs} & \multicolumn{6}{c}{Run Time\tablenotemark{a} (s)} & \multicolumn{2}{c}{Mean Relative Run Time} \\
	\cline{2-4} \cline{5-7} \cline{8-9}
	~ & \multicolumn{3}{c}{Real} & \multicolumn{3}{c}{Complex} & \colhead{Real} & \colhead{Complex} \\
	~ & \colhead{Default\tablenotemark{b}} & \colhead{RFI Excision\tablenotemark{c}} & \colhead{Windowing\tablenotemark{d}} & \colhead{Default} & \colhead{RFI Excision} & \colhead{Windowing} & ~ & ~
}

\startdata
  1 & 0.22 & 0.23 & 0.23 & 0.30 & 0.54 & 0.30 & 0.68 & 0.66 \\
  2 & 0.26 & 0.26 & 0.27 & 0.35 & 0.61 & 0.36 & 0.79 & 0.77 \\
  4 & 0.34 & 0.33 & 0.32 & 0.50 & 0.73 & 0.45 & 1.00 & 1.00 \\
  8 & 0.61 & 0.73 & 0.73 & 0.84 & 1.45 & 0.96 & 2.08 & 1.92\\
 16 & 1.28 & 1.33 & 1.36 & 1.82 & 2.95 & 1.72 & 4.00 & 3.81 \\
 32 & 2.48 & 2.84 & 2.61 & \nodata & \nodata & \nodata & 7.97 & \nodata \\
 64 & 4.84 & 5.06 & 5.18 & \nodata & \nodata & \nodata & 15.2 & \nodata \\
128 & 8.81 & 9.21 & 9.19 & \nodata & \nodata & \nodata & 27.4 & \nodata \\
\enddata

\tablenotetext{a}{Computations were run on an Intel i7 quadcore CPU clocked at 2.93 GHz with 16 GB of RAM.}
\tablenotetext{b}{Default signal processing with FFT channelization and integration.}
\tablenotetext{c}{RFI excision using time-domain blanking of samples above a specified power threshold.}
\tablenotetext{d}{Applying a Blackman-Harris window to the time domain data prior to performing the FFT.}

\end{deluxetable}

\clearpage
\begin{deluxetable}{cccccc}
\tablecolumns{6}
\tablewidth{0pc}

\tablecaption{Run time scaling of the {\bf lsl.correlator.fx.SpecMaster} function with number of samples\label{tab:IntegrationTime}}

\tablehead{
	\colhead{Samples} & \colhead{Integration Time\tablenotemark{b} (s)} & \multicolumn{2}{c}{Run Time\tablenotemark{a} (s)} & \multicolumn{2}{c}{Speedup} \\
	\cline{3-4}  \cline{5-6}
	& & \colhead{Real} & \colhead{Complex} & \colhead{Real} & \colhead{Complex}
}

\startdata
1.96 $\times$ 10$^5$ & 0.01 & 0.006 & 0.009 & 1.67 & 1.11 \\
9.80 $\times$ 10$^5$ & 0.05 & 0.030 & 0.041 & 1.67 & 1.22 \\
1.96 $\times$ 10$^6$ & 0.10 & 0.062 & 0.086 & 1.61 & 1.16 \\
9.80 $\times$ 10$^6$ & 0.50 & 0.300 & 0.455 & 1.67 & 1.10 \\
1.96 $\times$ 10$^7$ & 1.00 & 0.613 & 0.931 & 1.63 & 1.07 \\
2.94 $\times$ 10$^7$ & 1.50 & 0.925 & 1.306 & 1.62 & 1.15 \\
3.92 $\times$ 10$^7$ & 2.00 & 1.170 & 1.796 & 1.91 & 1.11 \\
\enddata

\tablenotetext{a}{Computations were run on an Intel i7 quadcore CPU clocked at 2.93 GHz with 16 GB of RAM.}
\tablenotetext{b}{Assuming a sample rate of 19.6 Msamples/s.}

\end{deluxetable}

\clearpage
\begin{figure}
	\plotone{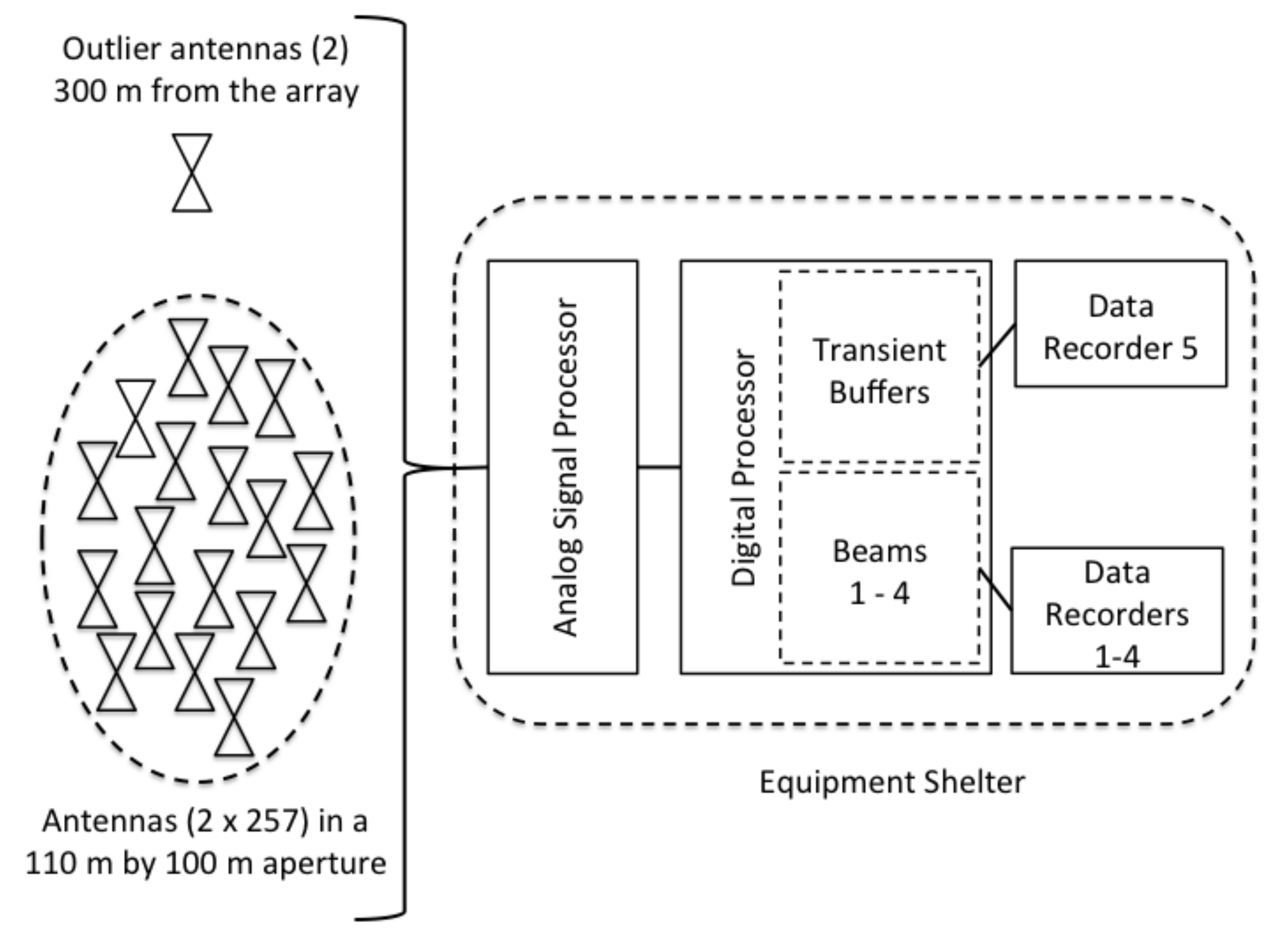}
	\caption{Block diagram of LWA1.  The LWA1 aperture is 110 m by 100 m and is populated by 257 stands, each supporting two crossed-polarization (north-south and east-west) active dipole antennas.  Also included at LWA1 is an outlier stand supporting two crossed-polarization antennas located 300m to the east of the array for calibration purposes.  The signals from each of the 516 dipoles enter the electronic shelter via coaxial cables and are filtered and amplified by the analog signal processor.  After filtering, the data are passed to the digital processor by Cat7 cables where the signals are digitized to 12-bits at 196 Msamples/s.  The digital processor is responsible for generating the transient buffer modes, TBW and TBN, and the four beams via the DRX mode.  The data generated by the digital processor are transferred to the data recorders via 10GbE cables where it is recorded.\label{fig:LWA1}}
\end{figure}

\clearpage
\begin{figure}
	\epsscale{0.75}
	\plotone{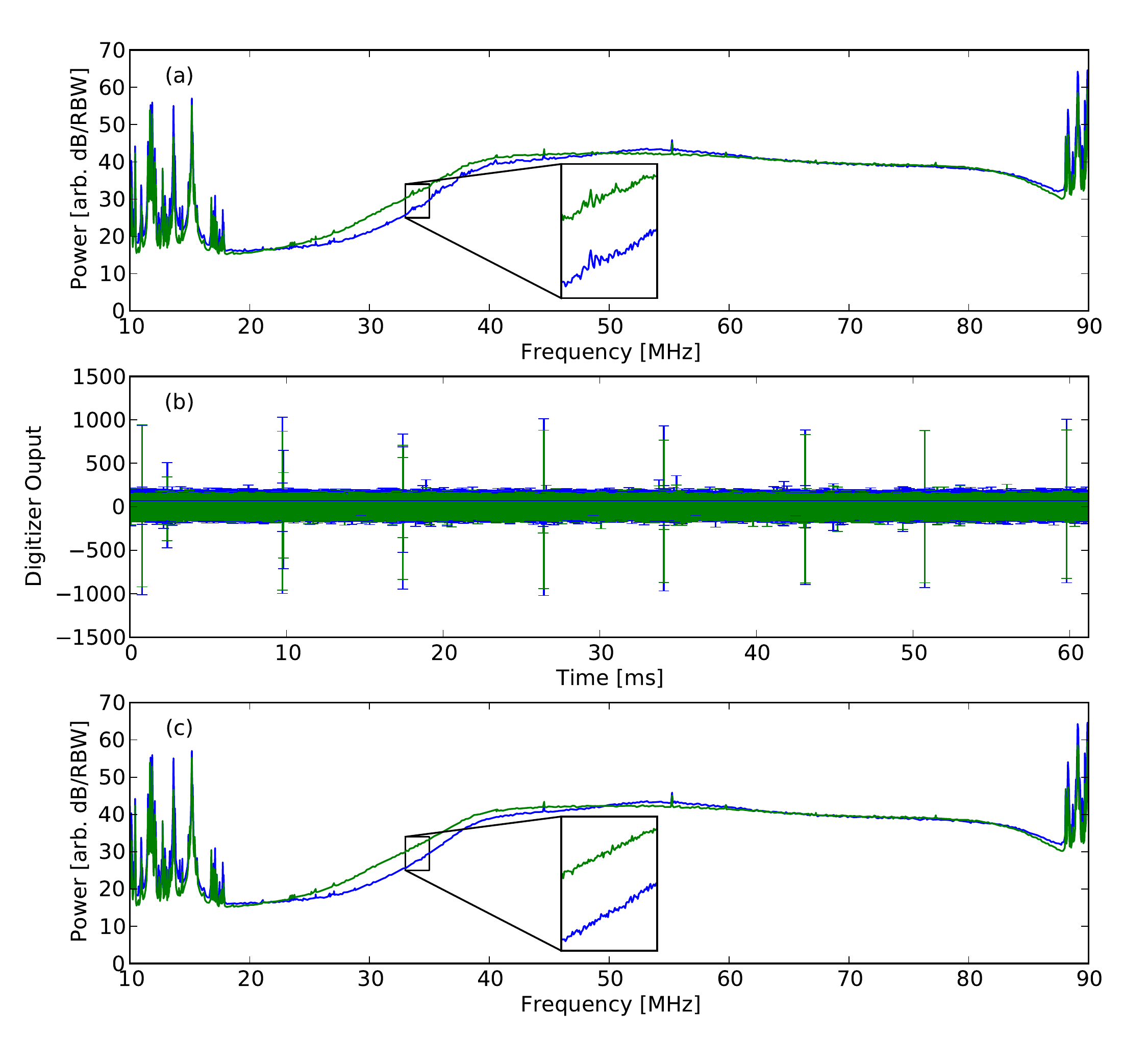}
	\caption{Panel (a) shows wideband transient buffer spectra of both the X (north-south; blue) and the Y (east-west; green) polarization dipole antennas located at stand 10 on April 18, 2012.  The spectrum is relatively clean in terms of RFI over the majority of the LWA1 band.  Notable exceptions to this are below 20 MHz where frequency bands are allocated for mobile transmitters and above 88 MHz in the FM band.  There is also broadband interference present in the spectra below 60 MHz.  The middle panel shows the time domain signature of this broadband interference as spikes in the digitizer values that last less that 0.1 $\mu$s and have a period of roughly 9 ms.  This is suggestive of micro-sparking on a power line near the LWA1 site.  Panel (c) shows the effect of dropping the FFT windows that contain the bursts on the integrated spectra.  This procedure is effective at removing the majority of the interference from the data.\label{fig:TBW}}
\end{figure}

\clearpage
\begin{figure}
	\epsscale{1.00}
	\plotone{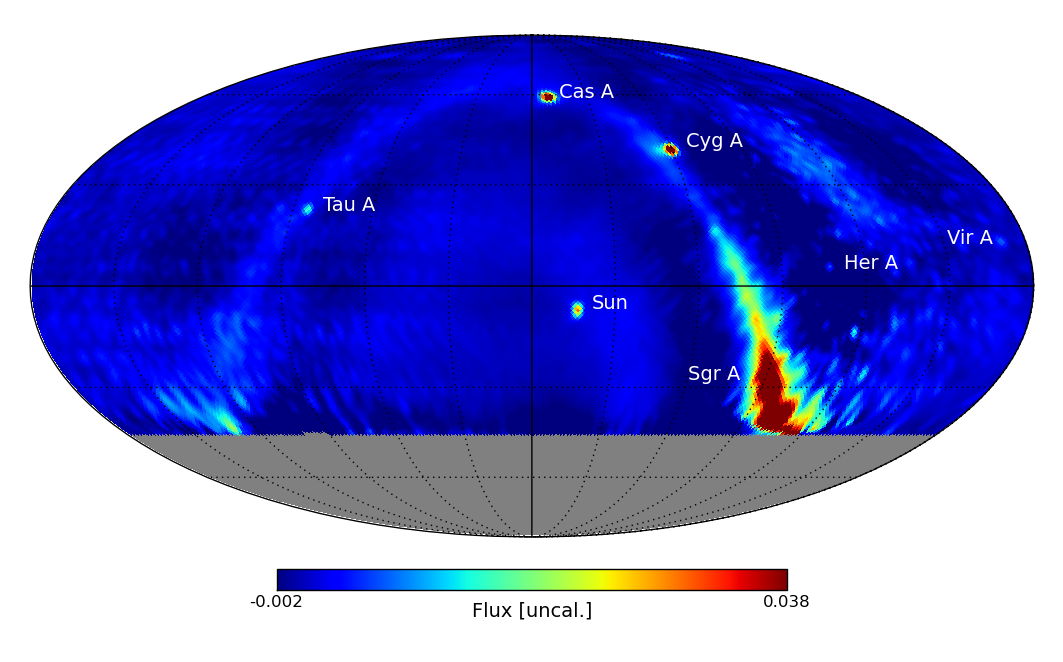}
	\caption{A Stokes I dirty image of the sky visible from LWA1 using narrowband transient buffer data at 74.03 MHz with 66 kHz of bandwidth.  The resolution of the map is approximately 1$^\circ$ and includes the sky above a declination of $\sim-$46$^\circ$. The solid lines denote the celestial equator (horizontal) and a right ascension of 0$^h$ (vertical).  Dotted lines mark the declination at 30$^\circ$ intervals and the right ascension at 2$^h$ intervals.  The sources Cyg A, Cas A, and the Sun as well as the Galactic plane are readily visible.  Also visible are side lobes of the Galactic plane which appear as areas of diffuse emission to the upper right and lower left of the plane.\label{fig:TBN}}
\end{figure}

\clearpage
\begin{figure}
	\epsscale{0.85}
	\plotone{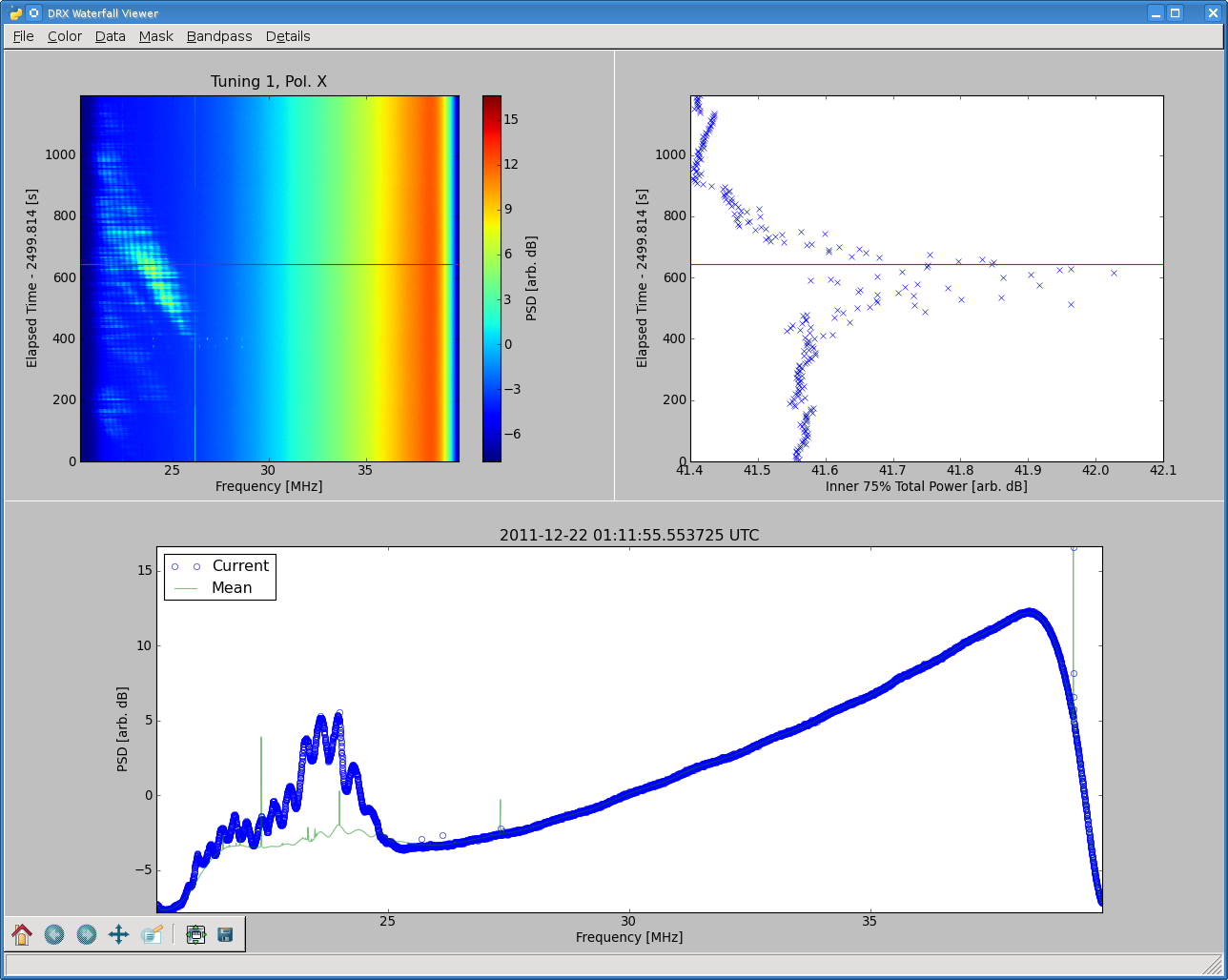}
	\caption{A waterfall plot created from beamformed observation of a Jupiter decametric burst from a 20 minute observation centered at 30 MHz displayed using the {\tt plotHDF.py} viewer provided with the {\tt Commissioning} extension (see \S\ref{sec:extensions}).  The observation was reduced to 8,192 channels with 5 second integrations using the {\tt hdfWaterfall.py} script.  The left-hand panel shows the waterfall with time increasing toward the top and frequency to the right.  The burst is likely a non-Io-A event and shows a complex structure in both time and frequency as well as Faraday lanes that drift from higher to lower frequencies over several minutes.  The right hand panel shows the total power for the inner 50\% of the bandpass as a function of time.  Finally, the spectrum at the bottom of the viewer provides a more detailed view of the burst at a particular instant.  The light green line denotes the integrated spectrum for the entire 20 minutes.\label{fig:DRX}}
\end{figure}

\end{document}